\documentclass{moriondNOTITLES}
\usepackage{xcolor}

\def\sinw{sin$\theta_{\textrm{eff}}^{\ell}$ }
\def\Afb{$A_{\textrm{fb}}$ }

\def\be{\begin{equation}}
\def\ee{\end{equation}}
\def\bea{\begin{eqnarray}}
\def\eea{\end{eqnarray}}

\newcommand{\Photo}{}

\begin{document}
\vspace*{4cm}
\title{QCD, Electroweak Physics, and Searches for Exotic Signatures in the Forward Region at LHCb}

\author{ Nathan Grieser, on behalf of the LHCb Collaboration }

\address{University of Cincinnati, Department of Physics\\
Cincinnati, Ohio, USA}

\maketitle\abstracts{
The LHCb experiment is a forward spectrometer that offers a unique phase-space coverage at the Large Hadron Collider (LHC).  Such a unique coverage offers the possibility to produce complementary and unique physics results in electroweak (EW), quantum chromodynamics (QCD), and searches for exotic signatures from beyond the Standard Model (BSM) physics.  These proceedings provide an exhibition of select results from the LHCb experiment in the fields of EW, QCD, and exotics. }

\section{Introduction}

The experimental discovery of the Higgs Boson in 2012 by the ATLAS ~\cite{ATLAS:2012yve} and CMS ~\cite{CMS:2012qbp} experiment provided a complete picture of the Standard Model (SM) particles.  Despite this astounding discovery, numerous measurements of the SM continue to be the highlight of the High Energy Physics (HEP) world.  The driving force for such precision measurements is to provide a probe for physics beyond the SM (BSM) ~\cite{Gaillard:1998ui}, where deviations from the expectation could provide an illumination on not yet understood physical phenomena.

The LHCb collaboration continues to produce significant precision measurements in the electroweak (EW) and hard Quantum Chromodynamics (QCD) fields to contribute to the larger global understanding of such topics.  Furthermore, a thriving exotics group compliments such measurements with direct searches for a variety of BSM models and archetypes.  This document shares a selection of recent results in the aforementioned fields and then briefly describes some exotic prospects for the Run 3 datataking period.  For a full list of publications from LHCb related to these fields, the reader is encouraged to visit Ref.~\cite{cern_alcm_analysis}.

\section{The LHCb Detector}

The LHCb detector~\cite{Alves:1129809}~\cite{LHCb:2014set} is a single-arm forward spectrometer at CERN’s Large Hadron Collider, designed for precision tracking and particle identification. It features multiple tracking detectors, a dipole magnet, and several particle identification systems including Cherenkov detectors, calorimeters, and muon chambers. Following the first major upgrade during Long Shutdown 2~\cite{LHCb:2023hlw}, the detector now supports higher luminosities and fully software-based event reconstruction, enabling enhanced exploration of exotic physics.

%The LHCb detector~\cite{Alves:1129809}~\cite{LHCb:2014set} is a single-arm forward spectrometer located at interaction point 8 of CERN's Large Hadron Collider. The detector includes a high-precision tracking system consisting of a silicon-strip vertex detector surrounding the $pp$ interaction region, a large-area silicon-strip detector located upstream of a dipole magnet with a bending power of about 4 T m, and three stations of silicon-strip detectors and straw drift tubes placed downstream of the magnet. Particle identification is achieved using information from two ring-imaging Cherenkov detectors, scintillating-pad and preshower detectors, as well as electromagnetic and hadronic calorimeters. Muons are identified by a system composed of alternating layers of iron and multiwire proportional chambers. The online event selection is performed by a trigger, which consists of a hardware stage, based on information from the calorimeter and muon systems, followed by a software stage, which applies a full event reconstruction.

%During the LHC's Long Shutdown 2, the LHCb experiment went through its first major upgrade period~\cite{LHCb:2023hlw}.  The tracking stations have all been upgraded, allowing similar precision performance as the Run 2 detector while taking increased data rates, significantly amplifying the EW and QCD physics program prospects.  Additionally, a revamped, fully-software based trigger selection allows for unique coverage of exotic state phase-spaces.

\section{Precision Electroweak Measurements}

Precision EW measurements serve as a probe of the SM and by definition~\cite{PhysRevLett.19.1264}  have sensitivity to new physics beyond the SM.  Equation \ref{eq:EWtheory} exemplifies the connection to the experimentally measurable free parameters in blue, and the higher-order, BSM sensitive, contributions in purple.

\begin{equation}
			m_{W}^{2}\left( 1 - \frac{m_{W}^{2}}{\textcolor{blue}{m_{Z}^{2}}}\right) = \frac{\pi\textcolor{blue}{\alpha}}{\sqrt{2}\textcolor{blue}{G_{\mu}}} \left( 1 + \textcolor{purple}{\Delta r}\right)
\label{eq:EWtheory}
\end{equation}

In recent years, LHCb has produced significant EW precision measurements, offering complimentary results to those produced by other LHC-based experiments due to the unique phase-space coverage, which gives differing sensitivities to PDF uncertainties than central measurements.  This section will briefly share the two most recent measurements: the measurement of \sinw and the measurement of the $Z$ boson mass.

\subsection{Measurement of \sinw with the LHCb Detector}

The weak mixing angle, $\theta_{W}$, is directly linked to the parameters of EW theory via Equation \ref{eq:sinw} and provides a relation between U(1) and SU(2) gauge couplings.

\begin{equation}
    sin\theta_{W} = \left( 1 - \frac{m_{W}^{2}}{m_{Z}^{2}} \right)
    \label{eq:sinw}
\end{equation}

\begin{figure}
\begin{minipage}{0.45\linewidth}
\centerline{\includegraphics[width=0.9\linewidth]{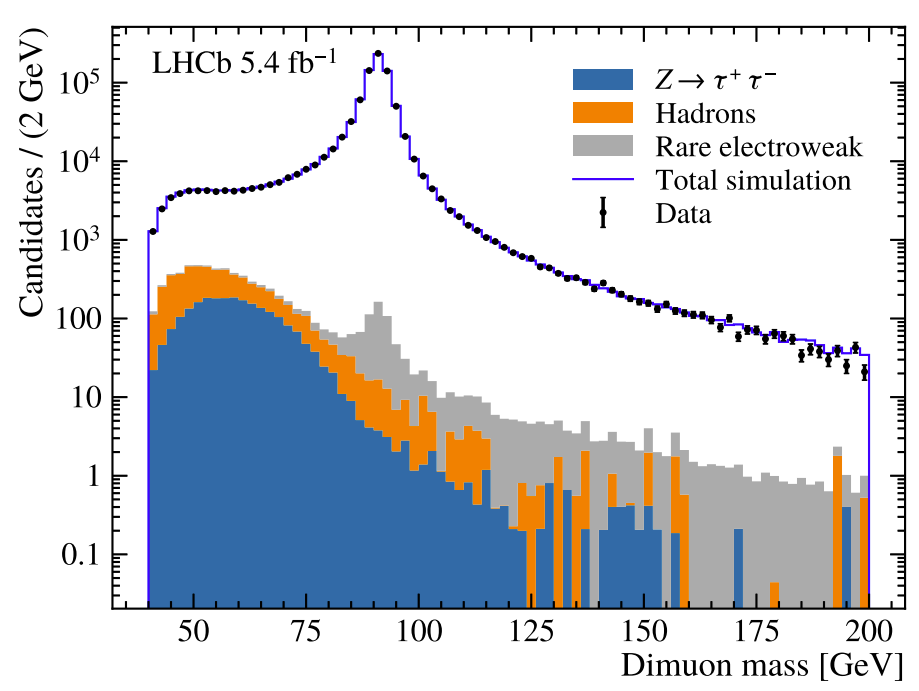}}
\end{minipage}
\hfill
\begin{minipage}{0.45\linewidth}
\centerline{\includegraphics[width=0.9\linewidth]{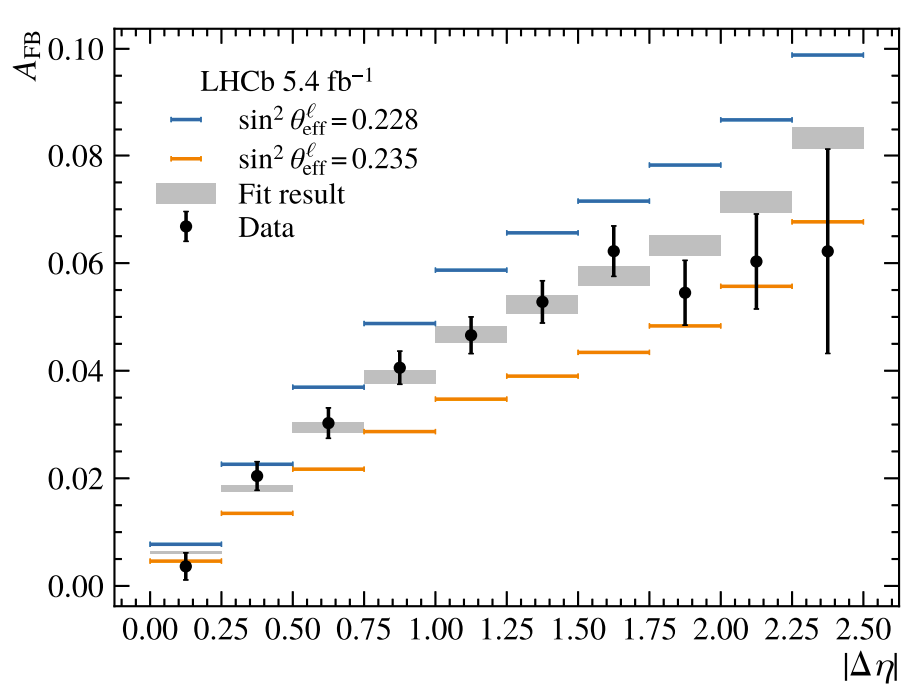}}
\end{minipage}
\caption{Signal and background contributions of the large window mass selection of the dimuon pair (left), and the results of the \Afb measurement differentially in bins of $|\Delta\eta|$ (right).  The sensitivity to different \sinw values is overlaid, highlighting the strength of differential selection.}
\label{fig:sinw}
\end{figure}

The recently published LHCb measurement~\cite{LHCb:2024ygc} leverages the connection of \sinw (which is directly correlated to sin$\theta_{W}$) to the forward-backward asymmetry (\Afb) of leptons produced in the decay of $Z$ bosons.  The muon pair final state provides an extremely clean sample of $Z$ boson decays, allowing the measurement of \Afb in bins of $|\Delta\eta|$ of the muon pair.  Visualization of the signal and background processes in the dimuon mass spectra, and the subsequent \sinw sensitivity of \Afb measured in bins of $|\Delta\eta|$ are provided in Figure~\ref{fig:sinw}.

The observed results of the effective weak mixing angle are sin$\theta_{\textrm{eff}}^{\ell}$ = 0.23152 $\pm$ 0.00044 (stat.) $\pm$ 0.00005 (syst.) $\pm$ 0.00022 (theory).  The uncertainties from proton PDFs are significantly smaller than in central experiments, offering exciting prospects for future high-luminosity LHC runs.

\subsection{Measurement of the Z Boson Mass}

Building on the measurement of \sinw , LHCb has made the first measurement of the Z boson mass ($m_{Z}$) at a $pp$ collider.  Similar to the \sinw measurement, the very clean signal environment of the dimuon mass spectrum allows for a precis measurement of $m_{Z}$.

The analysis selection is allowed to remain simple, necessitating precision gains to arise from accurate understanding of the LHCb detector and the produced dataset.  Numerous energy and momentum calibrations and corrections are employed, including the pseudo-mass method~\cite{Aaij_2024} which corrects data for curvature biases.

The observed value of $m_{Z}$ is found to be 91184.2$\pm$9.5 MeV, which is consistent with the SM expectation and similar precision to the global electroweak fit.

\begin{figure}
\begin{minipage}{0.45\linewidth}
\centerline{\includegraphics[width=0.9\linewidth]{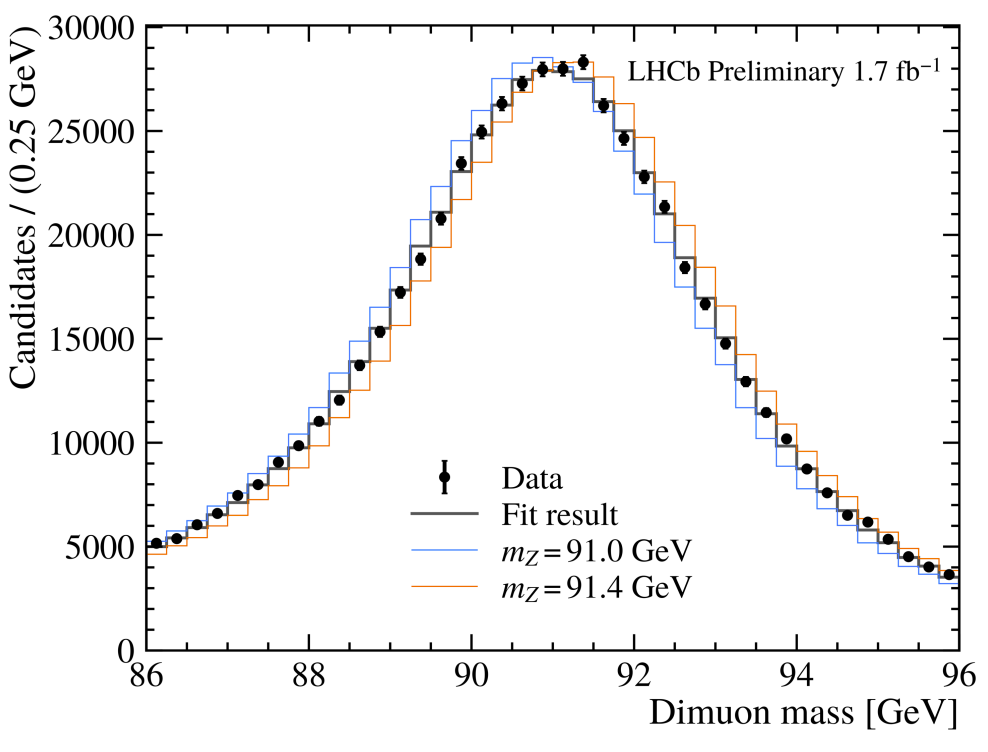}}
\end{minipage}
\hfill
\begin{minipage}{0.45\linewidth}
\centerline{\includegraphics[width=0.9\linewidth]{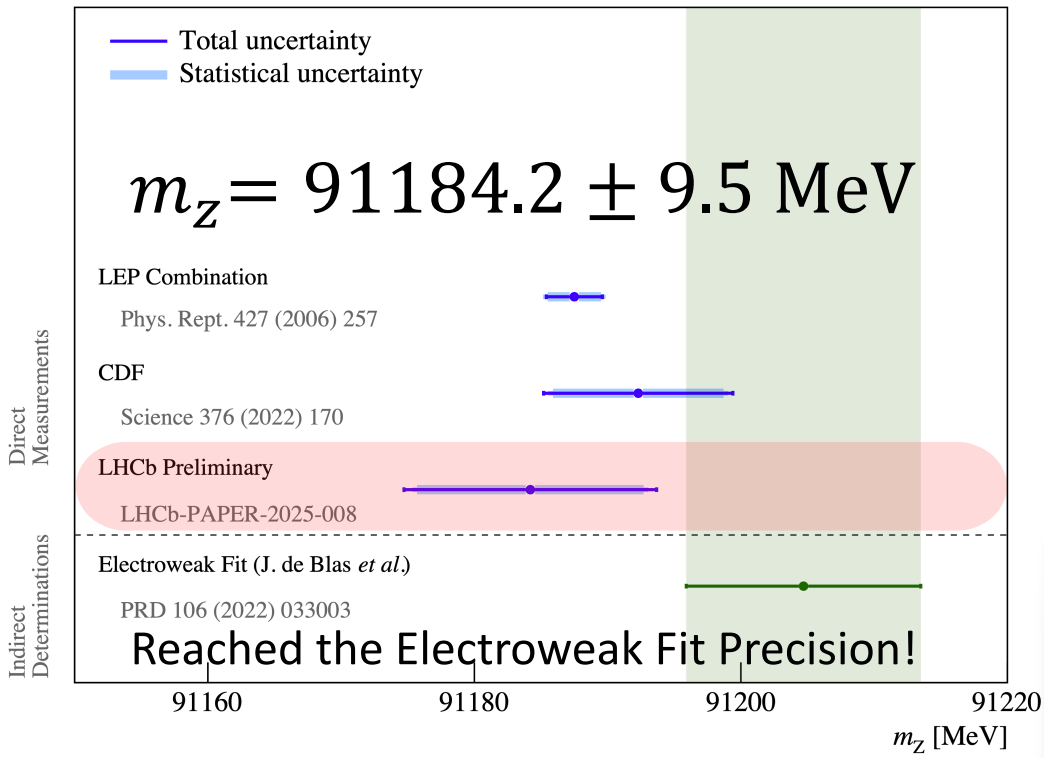}}
\end{minipage}
\caption{Fit of theoretical models to the LHCb 2016 dataset as function of the dimuon mass (left), and the measured value of $m_{Z}$ along with comparisons to other experimental measurements and the global electroweak fit (right).}
\label{fig:mZ}
\end{figure}

\section{Probing QCD With Lund Jet Plane Measurement}

For many years, QCD theory has held the expectation that there exists a suppression of collinear radiation around quarks during parton showering~\cite{Dokshitzer_1991}.  This phenomenon is referred to as the dead cone effect.  The dead cone effect can be observed experimentally with measurements of the Lund Jet Plane~\cite{PhysRevD.99.074027}, a 2-dimensional observable which details how particles inside a jet shower by angle and momentum.

Direct observation of the dead cone effect has been previously observed by the ALICE collaboration~\cite{Acharya2022} in charm decays.  The LHCb measurement expands on this observation by looking for a similar effect in $B$ decays.  A light jet enriched sample is collected by selecting jets recoiling from $Z$ boson decaying to a pair of muons.  A $B$ initiated jet sample is collected by selecting events with a fully reconstructed $B$ meson in the decay mode $B^{\pm} \rightarrow J/\psi (\rightarrow \mu^{+}\mu^{-}) K^{\pm}$.  To ensure a fair comparison between jets, a Winner-Take-All~\cite{Caletti2022} tag is utilized, requiring the heavy-flavor particle to always be the hardest at each decay node of the shower.  Lund jet planes are then populated for $k_{T}$ and $z$ variables using a declustering of jet constituents.

\begin{figure}
\begin{minipage}{0.49\linewidth}
\centerline{\includegraphics[width=0.9\linewidth]{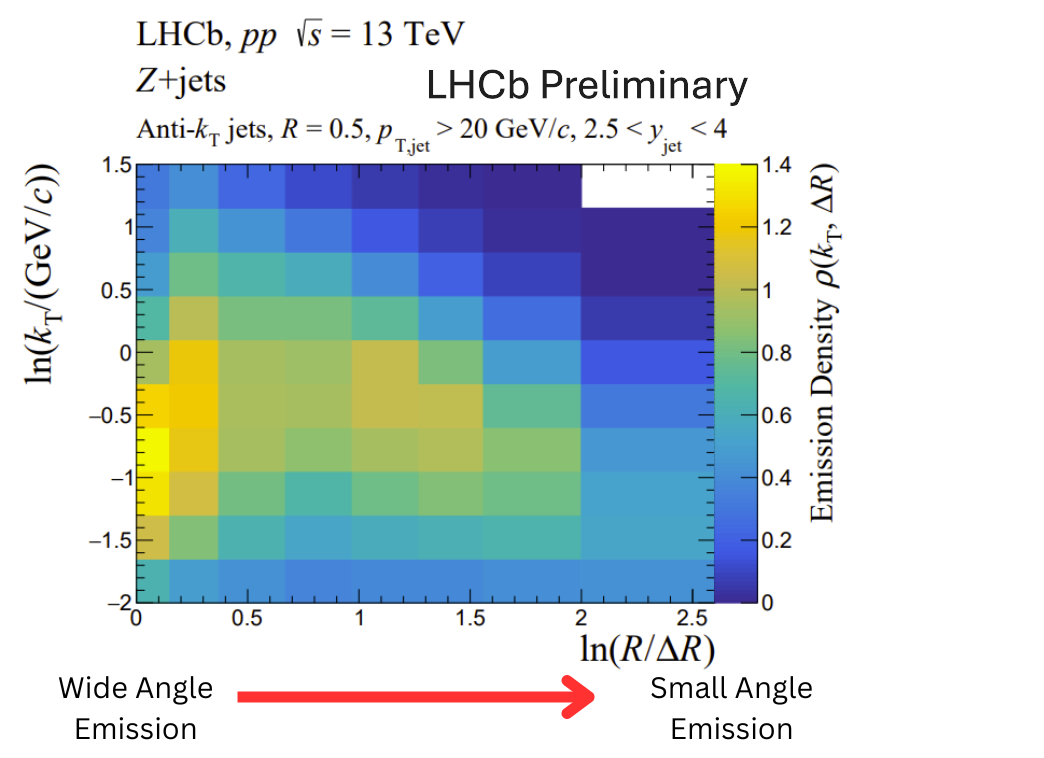}}
\end{minipage}
\hfill
\begin{minipage}{0.49\linewidth}
\centerline{\includegraphics[width=0.9\linewidth]{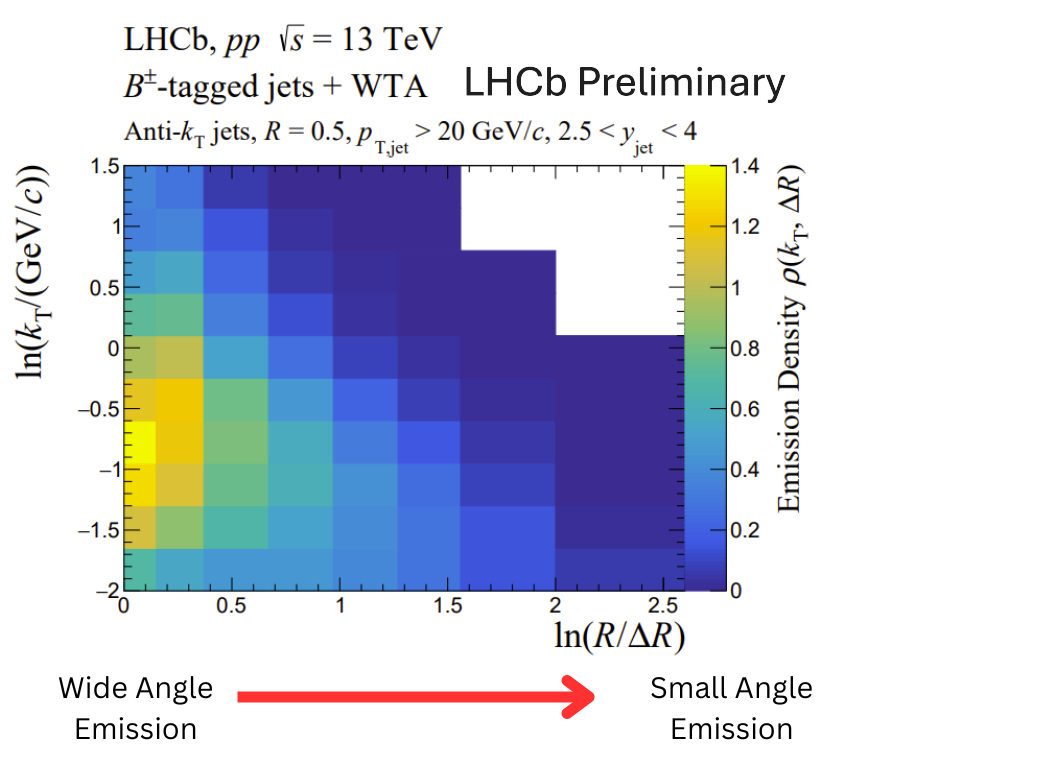}}
\end{minipage}
\caption{Lund Jet Planes in the $k_{T}$ variable for light jets (left) and $B$-tagged jets (right).  The suppression of small angle emissions for the higher mass jet initiators provides a direct observation of the dead cone effect.}
\label{fig:LJP}
\end{figure}

Measured Lund jet planes for $k_{T}$ are provided in Figure~\ref{fig:LJP}.  Suppression of small angle emissions is observed in the $B$-tagged jet sample in comparison to the light jet enriched sample, producing the first observation of the dead cone effect in $B$-initiated jets.

\section{Exotic Decays and Outlook for Run 3}

Significant EW and QCD measurements have been recently performed by the LHCb collaboration.  The significant increase in expected delivered integrated luminosity of Run 3 will significantly enhance these fields, alleviating the loss of precision due to limited sample sizes. Additionally, the exotic searches targeted by LHCb~\cite{Craik:2022riw}~\cite{LHCB-FIGURE-2025-001} with this sample have significant increases in sensitivity thanks to the fully software-based triggering and event reconstruction.  The coming years are expected to yield incredibly fruitful results from the LHCb electroweak, QCD, and exotics physics program.

\section*{Acknowledgments}

I would like to strongly thank the organizers of Moriond QCD for thought-provoking and well-ran environment to discuss many interesting physics topics.  Additionally, I would like to thank the LHCb collaboration for the opportunity to share the exciting new results shared in these proceedings.  Finally, I would like to acknowledge the funding support of the United States National Science Foundation which made travel for this opportunity possible.

\section*{References}

\addcontentsline{toc}{part}{References}
\bibliography{refs_Grieser}
\clearpage

\end{document}